\documentclass[12pt]{article} 
\usepackage{graphics} 
\usepackage{latexsym} 
\newcommand{\be}{\begin{equation}} 
\newcommand{\en}{\end{equation}} 
\newcommand{\ee}{\end{equation}}

\newcommand{\ra}{\rightarrow} 
\newcommand{\beq}{\begin{eqnarray}} 
\newcommand{\enq}{\end{eqnarray}} 
\newcommand{\cN}{{\cal{N}}} 
\newcommand{\p}{\partial} 
\begin{document} 
\bigskip \bigskip \bigskip 
\centerline{   \large \bf Warping and vacua of  (S)YM$_{2+1}$. } 
\bigskip 

\bigskip 
\centerline{\bf Iosif Bena and Aleksey Nudelman } 
\medskip 
\centerline{Department of Physics} 
\centerline{University of California} 
\centerline{Santa Barbara, CA  93106-9530 U.S.A.} 
\medskip 
\centerline{email:iosif and anudel@physics.ucsb.edu } 
\bigskip \bigskip 

\begin{abstract}         
We use dielectric branes to find non singular string 
theory duals of a perturbed 2+1 dimensional gauge theory living on D2 branes. By adding fermion masses we obtain theories with reduced supersymmetry. The Higgs vacua of the perturbed theory correspond to polarization of the D2 branes into D4 branes. The confining vacua correspond to polarization of the D2 branes into NS5 
branes. We consider different mass perturbations. Adding three equal masses preserves  $\cN=2$ supersymmetry.
In this case there are no confining vacua. By adding a fourth fermion mass we break all the supersymmetry, and find confining vacua. We also obtain duals for domain walls, condensates, baryon vertices, glueballs and flux tubes. We comment on the  Kahler potentials for the Higgs and confining phases.
In the course of the calculations we also find a nontrivial consistency check of the NS5 brane action in a D2 brane background.
\end {abstract} 

\section{ \large Introduction} 

In the context of the AdS/CFT correspondence \cite{juang}, 
Polchinski and Strassler \cite{joe} used   brane polarization  \cite{myers}, to  find  a string theory dual to a 
confining gauge theory in four 
dimensions.  They perturbed the  $\cN=4$ SYM theory living on N D3 branes by adding 
fermion masses 
and found a non singular dual spacetime which contained dielectric branes.  Giving masses to three out of  
the four  Weyl fermions broke supersymmetry to 
${\cal{N}}=1$, and the resulting gauge theory had a rich structure of 
vacua, domain walls, flux tubes, and instantons.  
 
Motivated by their approach we perturb the 2+1 dimensional SYM theory living on a set of coincident D2 
branes. We  add fermion mass terms and examine its supergravity/string theory  dual. Unlike the theory 
studied in \cite{joe}, this theory is not conformal, and there are different dual descriptions of the 
unperturbed theory
depending on the regime we are in \cite{imsy}. In particular, very  close to the branes the dilaton blows 
up, and IIA string theory ceases to provide a good description; consequently,  we find ourselves in M-theory.

Polarization of Dp branes into D(p+2) branes \cite{myers} can be understood in two ways. First, if we turn on a 
field which couples to a D(p+2) brane, a configuration with the Dp branes polarized into a wrapped 
D(p+2) brane is energetically favored. Alternatively, examining the superpotential in the weakly 
coupled theory, shows that the fields describing the coordinates of the Dp branes become 
noncommutative. The resulting configuration couples with a $p+3$ form (via a Born-Infeld term 
discovered in \cite{myers}) and has a nonzero transverse size; therefore it  can be interpreted as wrapped 
D(p+2) brane. 
 
For other types of polarizations (D3 branes into NS5 branes \cite{joe} or M2 branes into M5 
branes  the matrix model description is not well understood. The only direct description
is through energy considerations. Nevertheless, S-duality or the analysis of the symmetries of the M2 brane give indirect evidence for the second picture.
 
In \cite{joe}, adding  fermion mass perturbations results in the D3 branes being polarized into D5 or 
NS5 branes. If three Weyl fermions have equal masses, the $R$ symmetry is broken to $SO(3)$, and 
the five-branes have an $R^4 \times S^2$ geometry. If the three fermion masses are different, the shape 
of the five-branes  degenerates from a two-sphere into a two-ellipsoid.  
 
In \cite{iosif}, turning on four equal masses for the four complexified fermions  polarizes the  M2 branes 
into M5 branes of geometry $R^3 \times S^3$. If the four masses are not equal, the $SO(4)$ 
invariant configuration of the M5 branes degenerates into an ellipsoid.  
 
By perturbing the theory on D2 branes with fermion mass terms, we find configuration corresponding to 
D2 branes polarized into D4 or NS5 branes. The first polarizations is very similar to the D3 
$\rightarrow $ D5 polarization found in \cite{joe} and can be obtained from it via T-duality. The second 
polarization is reminiscent of the M2 $\rightarrow$ M5 polarization from \cite{iosif} and can be seen as 
coming from it under the compactification of the eleventh dimension.

The D4/D2 geometry is  smooth everywhere. For the NS5 brane the dilaton blows up  in the  throat region. However, the rate of growth of the NS5 brane dilaton in the throat region is much lower than that for the  D2 brane. Therefore the NS5/D2 geometry is less singular than that of D2 branes alone.
 
We may also study polarizations into NS5 branes with D4 brane charge, which should be similar to the 
oblique polarizations of \cite{joe}, but they are more complicated and will be the object of further study 
\cite{us-2}. 
 
This paper is organized as follows: 
In chapter 2 we perturb the IIA supergravity solution created by a large number $N$ of D2 branes, 
with the field corresponding  to fermion masses in the three  dimensional Yang -Mills. In chapter 3 we 
examine polarizations into D4 branes. We first consider a test D4 brane with large D2 brane charge in 
the perturbed geometry, and find that it gets   polarized into  $S^2$. We then extend the analysis to 
the full warped geometry created by shells of polarized D2 branes and  complete the analysis of 
the D2-D4 system by finding its near shell solution. 
 
In chapter  4 we study the polarization of D2 branes into NS5 branes. First we examine an SO(4) 
invariant fermion mass term, which does not preserve any supersymmetry and polarizes the D2 branes 
into NS5 branes wrapped on a 3-sphere.  Then we take one fermion mass to 0, which restores back 
$\cN=1$ and squashes the sphere into an SO(3) invariant ellipsoid. We  extend the analysis to the full 
warped geometry. Finally, in chapter 5 we  find string theory duals to domain walls, baryon vertices, flux tubes and 
glueballs.

\section{\large Perturbation of the bulk} 
 
According to the extension of the AdS/CFT duality to non conformal cases \cite{imsy}, the 2+1 SYM 
living on a set of $N$ D2 branes in a given regime is dual to type IIA string theory living in the near 
horizon geometry of these branes. 

The unperturbed space-time is given by:  
\beq 
ds^2&=&Z^{-1/2} \eta_{\mu \nu} dx^{\mu} dx^{\nu}+Z^{1/2} dx^{m} dx^{m}, \nonumber \\ 
e^{\phi}&=&g_s Z^{1/4}, \nonumber \\ 
C^0_3&=&-{1 \over g_s Z} dx^0 \wedge dx^1 \wedge dx^2, \ \ \ \ F_4^0 = d C_3^0, 
\label{metric}
\enq 
where $\mu,\nu = 0,1,2$, $i,j = 3,...,9$, $g_s$ is the string coupling and the metric is given in the string 
frame. When the D2 branes are coincident 
\be 
Z={R^5\over r^5}, \ \ \ \ R^5 = 6 \pi^2 N g_s {\alpha'}^{5/2}.
\label{2}
\en 

The supergravity description is valid when the curvature and the  dilaton 
are small. The Yang Mills coupling constant is related to the string coupling by $g_{YM}^2 = g_s /\sqrt{\alpha '}$. We are interested in  the boundary theory in the strongly coupled regime, which corresponds to 
small bulk curvature. As we go close to the D2 branes, the dilaton grows (the eleventh dimension opens up), 
and we have to use M-theory to describe the physics.

 The fermions on the two brane  transform in the $\bf 8$  of the $SO(7)$ R symmetry group \cite{seiberg}. 
The fermion bilinear and the mass matrix transform in a symmetric traceless representation {\bf 35}. This 
corresponds to antisymmetric three-tensors on the seven dimensional transverse space. Thus, we expect a fermion 
mass to correspond to a nonzero background for a three form field strength (which can only be NS-NS), and 
by Poincare\'e duality to a four form field strength (which is R-R).

The relevant IIA equations of motion and Bianchi identities are: 
\beq 
2 d(e^{-2 \phi} * H_3)=F_4 \wedge F_4, \nonumber \\ 
d(*F_4+B_2 \wedge F_4)=0,\nonumber \\ 
dF_4=0, \nonumber \\ 
dH_3=0. 
\label{eom} 
\enq 
 The background $F_4^0$ given by (\ref{metric}) and the first order 
perturbation $F_4^1$, which has components only in the transverse directions  are two 4-form field strengths.
Since both $F_4^1$ and $H_3$ have only transverse components, it is useful to express the ten
dimensional Hodge $*$ in terms of a 7-dimensional one obtained with a flat metric:
\beq 
{*} F_4=\frac{1}{Z} (*_7 F_4) \wedge  dx^0\wedge dx^1\wedge dx^2, \nonumber \\ 
{*} H_3=\frac{1}{Z^{1/2}}  (*_7H_3) \wedge dx^0\wedge dx^1\wedge dx^2. 
\label{stars}
\enq
By combining (\ref{eom}) with  (\ref{metric}) we obtain  
\beq 
d[1/Z(*_7 H_3+g_s F_4^{1})]=0, \nonumber \\ 
d[1/Z(*_7 g_s F_4^{1}+H_3)]=0. 
\label{pert-eom}
\enq 
We can see that  $F_4^1$ and $H_3$ are tensor spherical harmonics on the transverse space. We can 
express them in a basis for 3 and 4 tensors respectively, given in the Appendix:
\beq 
H_3 = r^p(c_1 T_3 + c_2 V_3) , \ \ \ \
F_4^1 =  r^q(c_3 T_4 + c_4 V_4).
\enq
Equation (\ref{pert-eom}) suggests the use of an ansatz:
\be
*_7T_3=T_4,
\en 
which implies
\be 
*_7V_4=T_3-V_3,
\en  
and its Hodge dual. Combining these with the relations in the Appendix, we obtain two solutions. The first 
one (which will interest us here) has a non normalizable mode 
\beq 
H_3=g_s\alpha/r^5 (3 T_3-5V_3), \nonumber \\ 
F^1_4=\alpha/r^5(4 T_4-5V_4), 
\label{nonnorm} 
\enq 
and a normalizable mode: 
\beq 
H_3=g_s \bar \alpha/r^7 (3 T_3-7V_3), \nonumber \\ 
F^1_4= \bar \alpha/r^7(4 T_4-7V_4). 
\label{norm}
\enq 
The other solution has the modes 
\beq 
H_3=-4 g_s \alpha/T_3,\nonumber \\ 
F^1_4=4\alpha/T_4 ,
\enq 
and 
\beq 
H_3=4g_s \alpha/r^{12}(-T_3 +4V_3),\nonumber \\ 
F^1_4=4\alpha/r^{12}(T_4-3V_4). 
\enq 
 These modes come from the  M-theory  anti self-dual and  
self-dual four form field strengths after the  eleventh dimension is compactified.

We now  explore the relation between the $H_3$ and $F^1_4$, which we turned on, and the $\cN=8$ 
gauge theory on the $N$ D2 branes. This theory has eight Majorana fermions transforming in the  $\bf 8$  
of the $SO(7)$ R symmetry group. The theory also has seven scalars in the fundamental representation of SO(7) and a vector.
In $\cN=2$ language,  six of these scalars can be combined with six fermions into three hypermultiplets, and one
scalar can be combined with the remaining two fermions and the vector into a vector multiplet. We see 
that giving masses to the three hypermultiplets  (six fermions) preserves $\cN=2$ supersymmetry. Nevertheless, if we give 
mass to the other two fermions we break supersymmetry completely.

In order to relate fermion masses to tensors on the transverse seven dimensional space it is convenient to 
group the eight fermions into four complex pairs, and to group six of the seven transverse directions into three 
complex pairs :
\beq
\Lambda^1&=&\lambda^1+i \lambda^2 \ ,  \ \  \Lambda^2 = \lambda^3 + i \lambda^4\ ,\ \ \Lambda^3 
=\lambda^5 + i\lambda^6\ ,\ \ \Lambda^4 = \lambda^7 + i \lambda^8 \ , \nonumber \\
z^1&=&x^3+ix^7\ ,\ \ \  z^2=x^4+ix^8\ ,\ \ \ z^3=x^5+ix^9.
\enq
In M-theory $x^6$ can be also grouped with the eleventh dimension into a complex pair, but this is not 
possible here.
Under a rotation $z^i \rightarrow e^{i \phi_i} z^i$ the fermions transform as :
\beq
\Lambda^1 &\rightarrow& e^{i(-\phi_1+\phi_2+\phi_3)/2 } \Lambda^1, \nonumber \\ 
\Lambda^2 &\rightarrow& e^{i(\phi_1-\phi_2-\phi_3)/2 } \Lambda^2, \nonumber \\ 
\Lambda^3 &\rightarrow& e^{i(\phi_1+\phi_2-\phi_3)/2 } \Lambda^3, \nonumber \\ 
\Lambda^4 &\rightarrow& e^{i(\phi_1+\phi_2+\phi_3)/2 } \Lambda^4.
\enq
If we perturb the Lagrangian with the traceless symmetric (in $\lambda_i$) combination 
\be
\Delta L = {\rm Re} (m_1 \Lambda_1^2+ m_2 \Lambda_2^2+ m_3 \Lambda_3^2+ m_4 \Lambda_4^2),
\ee
the corresponding bulk spherical harmonics transforming in the same way under SO(7) will be:
\beq
T_4& = & {\rm Re} \left(m_1 d\bar z^1 \wedge dz^2 \wedge dz^3 \wedge dx^6+ 
m_2 d z^1 \wedge d\bar z^2 \wedge d z^3 \wedge dx^6 \right. \nonumber \\ 
 &+& \left.m_3 dz^1 \wedge d  z^2 \wedge d\bar z^3 \wedge dx^6+ 
m_4d z^1 \wedge dz^2 \wedge dz^3 \wedge dx^6\right), \nonumber \\ 
T_3&= & {\rm Im} \left(m_1 d\bar z^1 \wedge dz^2 \wedge dz^3 + 
m_2 d z^1 \wedge d\bar z^2 \wedge d z^3  \right. \nonumber \\ 
 &+& \left. m_3 dz^1 \wedge d  z^2 \wedge d\bar z^3 + 
m_4d z^1 \wedge dz^2 \wedge dz^3 \right).
\label{fermass}
\enq
We chose the signs so that $T_4=*T_3$. The coefficient $\alpha$ from (\ref{nonnorm}) relates the 
fermion mass from the
gauge theory to the strength of the nonnormalizable mode, and will be determined in the next chapter.
Throughout this paper we are  keeping three masses equal: $m_1=m_2=m_3=m$. If $m_4=0$, we have 
$\cN=2$ supersymmetry, and SO(3) spherical symmetry. We can use this perturbation to study 
polarization of D2 branes into D4 branes. If $m_4=m$ we have SO(4) symmetry and no supersymmetry.
We can study polarization of D2 branes into $S^3$ wrapped NS5 branes. We can also study D2/NS5 
configurations for general $m_4$. In the limit $m_4 \rightarrow 0$ we recover supersymmetry.
\section{\large Polarization into D4 branes wrapped on $S^2$.} 

As we discussed in the introduction, there are two alternative ways to study polarization of D2 branes into 
D4 branes: using a matrix model description, or examining the potential for D4 branes with large D2 
brane charge. 
In order to find the self-interacting potential for $S^2$ wrapped D4 branes with large D2 brane charge, 
we first solve a simpler problem, and then, we use it to find the potential for a general 
configuration.

In the first subsection of this chapter we  explore a probe D4 brane  with D2 brane charge $n$, in 
the background created by a large number $N \gg n$ of coincident D2 branes (\ref{metric}) perturbed with 
(\ref{nonnorm}). We find a supersymmetric ground state at a nonzero radius. Then, we generalize to $N$ D2 branes 
distributed uniformly over a 2-sphere. We use this later generalization to find the full self-interacting 
potential of a spherical D4 brane with large D2 brane charge.

\subsection{\large D4 brane probes}

In 
a background formed by a large number of Dp branes
the action for a probe D(p+2) brane of world-volume $R^{p+1} \times S^2$ carrying Dp brane charge,  is 
\beq 
S&=&-\mu_{p+2}/g_s \int{d^{p+3} \xi Z^{(p-3)/4}\sqrt{-\det {G_{\parallel}} \det{(G_{\perp}+{\cal{ 
F}}_2)}}} \nonumber \\
&-& \mu_{p+2} \int{ (C_{p+3}+2 \pi \alpha ' {\cal {F}}_2 \wedge C_{p+1})}, 
\enq 
where $\mu,\nu =0 \ldots p$ and  $m,n=p+1 \ldots 9$. $G_{\perp}$ is a metric on a two sphere with 
$\det{G_{\perp}}=Z r^4\sin{ \theta}^2 $ and $G_{\parallel}$ is the metric in the  $R^{p+1}$ directions  
with $ \det{G_{\parallel}}=Z^{-(p+1)/2}$.  We have also defined 
\be 
2 \pi \alpha '{\cal{ F}}_2 \equiv 2 \pi \alpha 'F_2-B_2. 
\en 
The Dp brane charge of the D(p+2) brane is obtained by turning on a 2-form field strength along the 
$S^2$: $F_{\theta \phi}=(n/2) \sin \theta$ so that 
\be 
\int_{S^2} F_2=2\pi n,
\en 
and
\be 
F_{ab}F^{ab}=\frac{n^2}{2Zr^4}. 
\en 
This result does not depend on the particular form of $Z$. 
Specializing to the case of interest, $p=2$, we can expand  the integrand of the D4 action for  large  D2 
brane charge: 
\be 
\sqrt{\det(G_{\perp}+2 \pi \alpha' F)} \approx 2 \pi \alpha' \sqrt{\det F}+ 
\frac{\det G_{\perp}}{4 \pi \alpha' \sqrt{\det F}}. 
\en 
Using (\ref{2}), we see that the expansion is valid, provided 
\be 
4 \pi^2 F_{ab}F^{ab} \sim n^2 r/N g \sqrt{\alpha '} \gg 1 .
\en 
The leading term of the Born-Infeld action corresponds to the D2 brane charge, and is canceled by the 
second term in the Wess-Zumino action, as expected. 
The sub-leading term of the Born-Infeld action gives the potential per unit longitudinal volume
\be 
{-S_{BI} \over V}=\mu_4/g_s \int_{S^2} d^2 \xi Z^{-1/4}\frac{\sqrt{\det G_{\parallel}}\det 
G_{\perp}}{4 \pi \alpha' \sqrt{\det 
F}}=\frac{\mu_4 2 r^4}{g_s n \alpha '}, 
\label{lbi} 
\en 
where factors of $Z$ have canceled making the answer identical to the $D3/D5$ case. 
The integral of $C_5$ gives  a sub-leading term in the Wess-Zumino action. Since we 
know $F_4$ we can find $C_5$ by Poincare duality:
\be 
*F_4=-dC_5+H_3 \wedge C_3.
\label{C_5} 
\en 
The relative sign of $*F_4$ and $H_3 \wedge C_3 $ is found from the
equation of motion (\ref{eom}). The relative sign of $C_5$ and $H_3 \wedge C_3$ is determined 
by gauge invariance under the transformations 
\beq 
 \delta C_3=d\chi_2, \nonumber \\ 
 \delta C_5=-H_3. \wedge \chi_2 
\enq 
The component of $C_5$ which couples with the wrapped D4 brane, is the Poincare dual of $F_4^1$.
Using (\ref{stars}) and (\ref{nonnorm}) we find that up to a gauge choice 
\be 
C_5=-\frac{2 \alpha}{3R^5} dx^0 \wedge dx^1 \wedge dx^2 \wedge S_2. 
\en 
Calling $\zeta \equiv \alpha g_s /R^5$, we find the contribution of the Wess-Zumino term to be
\be 
{-S_{WZ} \over V} = -\frac{2 \mu_4 \zeta}{3 g_s} \int_{S^2}S_2. 
\label{cs5} 
\en 

By suitably choosing the plane of $S_2$ we can have a minimum of the effective potential 
away from 
$r=0$. The minimum of (\ref{cs5}) is when $S^2$ is in the 789 plane. A general SO(3) 
invariant brane configuration is obtained by rotating the sphere of radius $r$ in the 3-7, 4-8 and 5-9 
planes by the same angle $\theta$, and can be parametrized by $z=r e^{i \theta}$.
Therefore,
\be 
\int_{S^2}S_2=4 \pi\ \left ( 3m \  {\rm Im}( zz \bar z) + m_4 Im ( zzz) \right )
\en 
 Using (\ref{cs5}) we find the two main terms in the potential to be:
\be 
-{S_{WZ}+S_{BI} \over V}= \frac{2 \mu_4 |z|^4}{g_s n \alpha '}-\frac{8\pi \mu_4 m\zeta {\rm Im} zz 
\bar z}{g_s}. 
\label{actionin} 
\en 
Since we have supersymmetry in our theory, we expect the potential to be the square of the derivative of a 
superpotential, so we expect to have another term in (\ref{actionin}) proportional to $m^2 n$. This term 
comes from second order perturbation of the background metric, dilaton, and $C_{012}$ (via the second equation of (\ref{eom})). Since $C_{012} $ couples to $F_2$, this term is proportional to $m^2 n /g_s$.
Fortunately, supersymmetry allows us to read off this term by completing the square in (\ref{actionin}).
Using the convention in \cite{joe} we express the D2 brane collective coordinate $z$ in terms of the gauge 
theory scalar $\phi = z /2 \sqrt 2 \pi \alpha '$ and obtain:
\be 
{-S \over V} =    \frac{2 \mu_4 |z|^4}{g_s n \alpha '}-\frac{8\pi\zeta \mu_4 m {\rm Im} zz 
\bar z}{g_s} + \frac{8 \pi^2 \zeta^2 n \mu_4 m^2 |z|^2 \alpha'}{g_s} =
\frac{8}{g_{ym}^2 n} | \phi^2- i m n \phi  \zeta / \sqrt{2}|^2. 
\label{totalact}
\en 
We can compare this with the classical 3-dimensional $\cN=2$ Kahler potential and superpotential 
(which is identical to the one in \cite{joe} - equations (73) and (74)):
\beq 
K=\frac{n}{g_{ym}^2} \bar \Phi \Phi, \ \ \ \ W=\frac{mn}{2 g_{ym}^2} \Phi^2+\frac{2\sqrt{2} i }{3 
g_{ym}^2} \Phi^3. 
\enq 
They agree for  $\zeta=1/2$, which fixes the normalization of (\ref{nonnorm}).
There is a supersymmetric minimum at  
\be 
z =  i \pi m n \alpha '.
\label{D4minimum} 
\en 
Thus, a probe D4 brane with D2 charge $n$  has a supersymmetric ground state in the 789 plane at 
radius $|z|$. As one may notice, the potential (\ref{totalact}) does not depend on $x_6$. Therefore, the D4 sphere
can be continuously moved along $x_6$. This modulus can in principle be lifted by quantum corrections, and deserves 
a more complete investigation. 

\subsection{\large The full problem}

In this section we  find the full potential for a D4 brane with D2 brane charge $N$ in the geometry 
created by itself. Since this is a self interaction problem, we have to find the potential by bringing D4 
shells with D2 charge from infinity in the background created by polarized D2 branes. The only difference 
from the previous chapter is the new $Z$. Calling $r_0$ the radius of the 2-sphere in which the 
branes are polarized, $w$ the radius in the polarization plane and $y$ the radius in the transverse 
directions we obtain:
\be 
Z=\frac{R^5}{6 r_0 w}\left (\frac{1}{(y^2+(w-r_0)^2)^{3/2}}-\frac{1}{(y^2+(w+r_0)^2)^{3/2}} \right ).
\label{Z} 
\en 
Far away from the branes this reduces to (\ref{2}). One might expect the calculation to be much
harder, but using a clever trick from \cite{joe}  we do not have to compute anything. Equations 
(\ref{pert-eom}) imply that
\be 
d [Z^{-1}(H_3+g_s *_7F_4)] = d *_7   [Z^{-1}(H_3+g_s *_7F_4)] = 0;
\en 
therefore, $Z^{-1}(H_3+g_s *_7 F_4)$ is constant and equal to its value at $\infty$ (which is given by the 
gauge theory). In particular,  (\ref{C_5}) implies  that $C_5$ remains unchanged;  thus, the Wess-
Zumino term is the same. Moreover, we have seen  that the Born-Infeld term (\ref{lbi}) does not depend on 
$Z$ either. The third term of the potential is related to the first two by supersymmetry, and therefore it is
also invariant. Thus, there is no change in the potential of a probe brane if the D2 branes 
creating the geometry are spread out. Because of the above, the total potential is the same as the probe 
potential.

A general ground state is given by D2 branes polarized into several shells of D4 branes, with D2 brane 
charges $n_i$.  As we have discussed, the potential felt by a D4 shell depends only on its radius, and not on $x^6$ or on the position of the other shells. Thus, the shells may be located at different positions along $x^6$. This moduli space can in principle be destroyed by quantum corrections, and we can also destroy it classically by giving a mass to $x^6$.

The action is a sum of the actions (\ref{totalact}), each with its own $n_i$.
The results in this section are similar to the ones obtained for the  D3-D5 polarization in  \cite{joe}.
Indeed, by performing a T -duality along $x^6$  we can obtain all the D3-D5 configurations from D2-D4 configurations.

\subsection{The near shell solution } 
In finding the equilibrium configuration for the D2-D4 system we assumed that 
$ F_2$ is dominant in the Born-Infeld term. This approximation breaks down close to the D4 brane shell,
where we need to account for the deformation of the metric due to the D4 charge.  Since very close to the shell 
we can approximate the shell to be flat, we can match the metric (\ref{metric},\ref{Z}) with that of $p$ 
flat  D4 branes with magnetic flux \cite{oz}. This metric is:

\beq 
ds^2_{string}&=&\frac{ u^{3/2}}{p a^{3/2} \pi g_s \sqrt{\alpha'}} \left [ \eta_{\mu \nu} dx^{\mu} 
dx^{\nu}+h(d \tilde{x^3}d \tilde{x^3}+d \tilde{x^4}d \tilde{x^4}) \right ] + \nonumber \\ 
&+& \frac{\sqrt{\alpha'} p a^{3/2}  \pi g_s}{ u^{3/2}} (du^2+u^2 d \Omega_4)\ ,\nonumber \\ 
e^{2 \phi} &=& \frac{g_s^2 u^{3/2} a^3R^{3/2}}{ (1+a^3u^3)}\ ,\ \ \ \  h=\frac{1}{1+a^3 u^3}. 
\label{d4near}
\enq 
The warped metric near the shell, at a point which we can fix without loss of generality to  
$(w_1,w_2,w_3)=(0,0,r_0)$ is:
\be 
ds^2_{string}=\frac{\sqrt{6} r_0 \rho^{3/2}}{R'^{5/2}}\eta_{\mu \nu} dx^{\mu} 
dx^{\nu}+\frac{R'^{5/2}}{\sqrt{6} r_0 \rho^{3/2}}(dw \cdot dw+dy \cdot dy), 
\label{d4far}
\en 
where $\rho^2=(w_3-r_0)^2+y^2$ and $R^{3/2}=\pi g_s \sqrt{\alpha '} p a^{3/2}$ ,  $R'^5=6 \pi^2 g_s n 
\alpha '^{5/2}$. 
For large $au$ the metric (\ref{d4near})  matches  (\ref{d4far}) provided
\beq 
u=\rho \alpha'^{1/3},\ \ \ \   a^3=\frac{n \alpha'^{5/2}}{g_s r_0^2 \pi ^2}, \ \ \ \  \tilde{x}^2=\frac{\pi g_s^{3/2} \sqrt{n} \alpha'^6 n^2}{p^2 r_0^4} w^2.
\enq  

Let us briefly describe the ranges of validity of the supergravity solutions.
There are two regions of interest.
The D4 branes can be considered approximately flat when we are very close to them. This assumption is valid 
when $\rho_c \ll r_0$.
On the other hand, the supergravity solution (\ref{d4far}) is valid when the dilaton is small, which is 
always the  case near the crossover distance, $au=1$. This gives  
\be 
\rho_c \sim(g_s p^2/n)^{1/3}\sqrt{\alpha'}.
\en 
Thus, the near shell approximation is consistent because $\rho_c$ is much smaller than the size of the sphere $r_0$. The supergravity approximation is valid if the size of the transverse sphere $ R^{3/2}/\sqrt{a}$ is not too small. In the conformal case this quantity is independent of $r_0$.
In our case the radius of the transverse 4-sphere is
\be
\left ( \frac{\pi g_s^2 \alpha'^2 p}{nm^2} \right )^{1/6}.
\en 
Therefore, we can trust the  supergravity  approximation for small $m$ and /or large $p$.

\section{\large NS5 brane probes} 
 
As we have seen, a background with nonzero $H_3$ and $C_5$ supports a D4 brane at nonzero radius. 
Nevertheless, the field equations of motion couple $H_3$ with $F_4$, so we expect the background to 
support an NS5 brane at nonzero radius as well. The type IIA $NS_5$ is quite a mysterious object with a 
self-dual 2-form field living on it, and its action has been constructed only recently \cite{sorokin}.  
 
We first investigate an NS5 brane with the geometry $R^3\times S^3$, and with a large number $n$ 
of D2 branes dissolved in it, in the background created by an even larger number $N$ of D2 branes.
Turning on a D2 brane charge $n$ is done by turning on a 3-form field strength 
along the $S^3$. This has to be done carefully because the field strength is self dual. 
As we shall see, self duality in this case is nontrivial.

\subsection{\large The NS5 brane action} 
 
The NS5 brane action in a background with no RR 1-form flux is very similar to the M5 brane action.

There are two formulations of the M5 brane action. The first is a manifestly covariant formulation (first obtained by Pasti 
Sorokin and Tonin \cite{pst}) which involves an auxiliary fields. The second formulation (first obtained by Perry and Schwarz in \cite{ps}, and generalized for general gravitational backgrounds in \cite{schwarz}) has no auxiliary fields but only 5-dimensional manifest covariance . 
This action is more amenable to explicit calculations, and  can be obtained from the first one by gauge fixing.

The formulation of the NS5 brane action in \cite{sorokin} is very similar to the  PST \cite{pst} M5 brane action, 
in that it contains an auxiliary field, which imposes the self-duality 
of the 2-form field when integrated out. Moreover, in order to do explicit computations, the   NS5 brane action 
has to be gauge fixed. In the absence of background RR one-forms, the gauge fixing of NS5 brane action is similar to the reduction of the PST action to the PS action \cite{costin,pst} in M-theory.
 
For zero RR one-form flux and for the dilaton background (\ref{metric}) the NS5 brane action becomes: 
\beq 
S&=&-\mu_{5} \int d^6 \xi [L_{BI}+L_2+L_{WZ}],  \nonumber \\  
L_{BI}& =& g_s^{-2}Z^{-1/2} \sqrt{-\det{(G_{mn}+i g_s Z^{1/4} D_{mn})}}, \nonumber \\  
L_2 &=& \sqrt{-G} {1 \over 4 \p_r a \p^r a} \p_m a (*D)^{mnp} D_{npq}\p^q a  ,  \nonumber \\ 
L_{WZ}&=& [B_6-{1 \over 2} F_3 \wedge C_3],  
\enq 
where $B_6$ and $C_3$ are the pullbacks of the IIA forms, $D_3\equiv F_3-C_3$, and $ D_{mn} = 
(*D)_{mnp} {\p^p a \over \sqrt{ \p_r a \p^r a }} $. The first and the last terms look like  Born-Infeld and  
 Wess-Zumino terms respectively.

If we fix the gauge to $a=x^2$, $b_{2m}=0 $, we recover an action with reduced explicit Lorentz 
invariance, similar to the PS action: 
\beq 
S = -\mu_{5} \int d^6 \xi \ g_s^{-2} Z^{-1/2} \sqrt{-\det{(G_{mn}+i g_s Z^{1/4} D_{mn})}}+  
\nonumber \\ 
 + {1 \over 4} \sqrt{-G \over G_{22}} D^{mn} D_{mn2} + L_{WZ}, \hspace{2cm} 
\enq 
where $D_2$ has been changed to:  
\beq 
D^{mn}=\frac{\sqrt{G_{22}}}{3! \sqrt{-G}} \epsilon^{2mnpqr} D_{pqr}. 
\enq 
The equations of motion are the same as in \cite{schwarz}, with an extra $g_s Z^{1/4}$ multiplying 
$1/\sqrt{-G/G_{22}}$. When only three-form fields in orthogonal directions are turned on, the 
equations simplify to give: 
\beq 
D_{mn2} = \p_2 B_{mn}-C_{2mn}=  { \sqrt{G_{22}} D_{mn} \over \sqrt{1 + g_s ^2Z^{1/2}D_{mnp} 
D^{mnp} } }. \label{ns5eom}
\enq 
 Under this assumption also  
\beq 
\sqrt{-\det{(G_{mn}+i g_s Z^{1/4} D_{mn})}} = \sqrt{-G}\sqrt{1+  g_s^2 Z^{1/2}D_{mnp} D^{mnp}} .
\enq 
 
\subsection{\large The NS5 brane probe}

We first consider a situation with four equal fermion masses, where we have SO(4) symmetry, but no 
supersymmetry. In the absence of supersymmetry the masses of the seven scalars are not constrained.

We study a test configuration in which a large number of D2 branes is extended in the  012 
directions; a probe NS5  brane has three directions wrapped on an $S^3$ in the $3456$ plane, and the 
other three are parallel to the D2 branes. We call $\hat \epsilon_{ijkl}$ the numerical antisymmetric tensor 
restricted to the $3456$ plane. We also give the NS5 brane a D2 brane charge $n$,  by turning on a 3-
form 
field strength along $S^3$: 
\be 
F_3 = {A\over r^4 3!} \hat \epsilon_{ijkl}\ x^i\ dx^j \wedge  dx^k  \wedge dx^l =  {A } \sin^2 
\theta 
\sin \theta_1\ d \theta \wedge d \theta_1 \wedge d \phi ,
\label{f3}
\en 
where 
\be
A = 4 \pi n (\alpha')^{3/2}, 
\ee
and $ \theta$ , $\theta_1$ and $\phi$ are the angles on the three-sphere.
For now we assume $n < N$, so that the effect of the brane probe on the background is negligible. 
Equation (\ref{fermass}) also gives us  $T_{3456}= 4m $; thus,  
\be 
C_3^1={Z \over 2g_s} S_3 = {4m \over 2g_s 3!}   \left({R \over r}\right)^5  \hat \epsilon_{ijkl} x^i dx^j 
\wedge  dx^k  \wedge dx^l =  { 2m R^5 \over g_s r} \sin^2 \theta \sin \theta_1\ d \theta \wedge d \theta_1 
\wedge d \phi,
\label{c3}
\en 
$C_3^0$ is given by (\ref{metric}). 
The 6 form dual of the NS-NS 2 form can be found using (\ref{eom}) from which we obtain 
\be 
d B_6 - {1\over 2} C_3 \wedge F_4 =  F_7 = e^{-2 \Phi} * H_3.  
\en 
Using (\ref{metric}), (\ref{nonnorm}) and the relations in the Appendix, we obtain  
\be
dB_6 =[ Z^{-1}({*_7 H_3 \over g_s^2} +{F^1_4 \over g_s})  + {1 \over 2}d( C_3^1 \wedge C_3^0)]  \wedge  dx^0  \wedge dx^1 
\wedge 
dx^2 = 0.
\label{B6}
\ee
Thus, the first term in the Wess-Zumino action gives no contribution, like in the M-theory case 
\cite{iosif}. This is to be contrasted with the non-vanishing $dC_5$ in the D4 brane case and the 
non-vanishing $dC_6$ and $dB_6$ in \cite{joe}.

Since we have spherical symmetry, the value of the action is the same at every point on the 3-sphere. 
At the point $x^6=r$: 
\be
D_{\perp} = F_{345}-C_{345} = - \left[{{A \over r^3} - {2mR^5 \over g_s r^4}  }\right] = - {A 
\over 
r^3}  - C_{345}.
\ee 
We are interested in the limit when the D2 brane charge of the NS5 brane is bigger than its NS5 brane 
charge. This means $ n^2 g_s \sqrt{\alpha'} \gg r N$. In this 
limit, 
$F_{345}$ will be dominant in the Born-Infeld term.
Using the equation of motion (\ref{ns5eom}) we obtain:  
\be
\p_2 B_{01} - C_{201}=  {-  \sqrt{-G_{\parallel} G_{\perp}^{-1}}  D_{\perp} \over  
\sqrt{1+ g_s^2  Z^{1/2}  G_{\perp}^{-1} D_{\perp}^2 } } = {- D_{\perp} Z^{-3/ 2}\over  
\sqrt{1+ g_s^2  D_{\perp}^2 Z^{-1}  } }. \label{duality}
\ee

To find $F_{012}$ we observe that for large D2 brane charge the 
right hand side of (\ref{duality}) gives exactly the background value of $-C_{201}$, so the contribution of 
$F_{012}$ to the Wess-Zumino term is negligible.

We have discovered a very interesting fact. The equations of 
motion of an NS5 brane with large D2 brane charge in the geometry (\ref{metric}) give rise to a ``dual'' 3 
form equal to the 3-form field of this geometry. 
The bulk $C_3$ created by the dissolved 
branes is the same as the one obtained via the NS5 brane equation of motion. 

This is an interesting connection between the IIA supergravity  and the NS5 brane equations of motion which deserves further study. Note that this result is independent of $G_{\perp}$, and thus, it holds for any NS5 brane shape at any warp factor. This phenomenon is very similar to the one observed in the case of the M-theory 5-brane \cite{iosif}, and is probably a general feature of actions with a self-dual field.

We now proceed to the calculation of the potential per unit longitudinal volume felt by the NS5 brane 
in this geometry. The dominant parts of the potential are:
\beq
{-S_{BI} \over 2 \pi^2 \mu_5 V}  &=& r^3 {Z^{-1/2} \over g_s^2} \sqrt{-G_{\parallel} G_{\perp}} \sqrt{ 
1 + g_s^2 D_{\perp}^2 Z^{1/2} } \approx  {A \over g_s Z} +  { r^6  \over 2 g_s^3 A } + { r^3 C_{345} 
\over g_s Z}, \nonumber \\
{-S_{mixed} \over 2 \pi^2 \mu_5 V}  &=&  {-r^3 \over 2} { D_{\perp}^2 Z^{-3/2} \over  \sqrt{1 + g_s^2 
D_{\perp}^2 Z^{-1} } } \approx -{A  \over 2 g_s Z} + { r^6  \over 4 g_s^3 A } - { r^3 C_{345} \over 
2g_s Z},   \nonumber \\
{-S_{WZ}\over 2 \pi^2 \mu_5 V}   &=& -{r^3 \over2} [- 2 B^6_{012345}+ C_{012}F_{345} - 
F_{012}C_{345}] \approx  -{A  \over 2 g_s Z}.
\label{NS5-potential}
\enq
where the approximation is valid for $n^2  g_s \sqrt{\alpha '} \gg r  N$. We have also included the $B^6$ 
component which in the probe brane case is zero, but might be non-vanishing for a general warped geometry.
The first terms represent the interaction of the dissolved D2 branes. They are 
proportional to $g_s^{-1}$, as D brane action terms are, and they cancel because parallel D2 branes do 
not interact. The terms proportional to $r^6$ are the gravitational energy of the NS5 brane, and they are 
attractive. The terms proportional to $C^3$ are repelling. The effect of those two terms is to favor a ground 
state at nonzero radius.

Nevertheless, as in the D4 brane case, there can be another term in the potential of the same relevance as the first two. We expect a 
second order correction to $C_{012}$, via the second equation in (\ref{eom}) as well as second order corrections to the metric and dilaton. Using the first order fields 
(\ref{nonnorm}) and remembering that $C_{012}$ couples with $F_{345}$, we can see that the extra 
term is proportional to $m^2 A  r^2 /g_s$, and is of the same order as the first two terms. The potential is 
therefore:
\be
{-S \over 2 \pi^2 \mu_5 V} = {3 r^6 \over 4 g_s^3 A} - {m \over g_s^2} r^4  +{c A m^2 r^2 \over g_s},
\label{52}
\ee
where $c$ is a constant. Determining $c$ is not trivial because of the second order corrections to the 
metric and the dilaton. In the first part of this paper, as well as 
in \cite{joe, iosif} $c$ is computed  by invoking supersymmetry. 

Unfortunately, with four fermion masses, there is no supersymmetry. If we give up one fermion mass, and thus give 
up SO(4) invariance for SO(3) invariance, we can still use supersymmetry to determine the value of $c$. This will 
be the subject of section 4.4. 

We can also find $c$ by noticing that the last term of (\ref{52}) represents a mass for the scalars. This will be
the subject of chapter 4.5.
 
\subsection{\large NS5 brane warping}

In the general case the D2 branes are polarized into a sphere with NS5 brane charge. We try to 
find
the potential for this configuration. Since we are doing a `self-force' problem, we have to find the potential 
by bringing small D2 brane spheres one by one from infinity. In classical electromagnetism this gives the 
familiar factor of 1/2 for the self energy of a charged object.

We first consider a test NS5 brane shell with D2 charge in the geometry created by a large number of 
polarized D2 branes. The only difference from the previous chapter is the new $Z$.  We can 
use an argument similar to the one we have used for D4 branes to show that the potential does not change. The 
equations (\ref{pert-eom}) imply that $*_7H_3+g_s F_4^1$ is harmonic. Its behavior at infinity is given by 
the boundary theory, and is the same as for trivial $Z$. Therefore, 
\be
*_7H_3 + g_s F_4^1 = 2 Z T_4.
\ee
Using (\ref{B6}) and (\ref{NS5-potential}) we  see that the particular combination which enters the 
NS5 brane action, $B_6- {1\over 2} C_3^1 \wedge C_3^0$, is not changed. This ``conspiracy'' is similar to 
the one in \cite{iosif}. The $C^1 \wedge C^0$ term (coming from the BI and the mixed action) appears in 
the effective NS5 brane action and in the first order bulk equation of motion with the same coefficient 
relative to $B_6$. Thus, in a general $Z$ background $B_6$ and $C_3^1$ may change, but an NS5 brane 
with large D2 charge does not feel that.

The large terms in (\ref{NS5-potential}) cancel as usual, and the terms proportional to $r^6$ do not 
depend on $Z$. The last term may change, but in the configurations we care about it is fixed by 
supersymmetry, so it does not depend on $Z$ either. Since the potential felt by a probe brane does not 
depend on the distribution of branes, the total potential is the same as the probe potential.

One may also worry about the effect of the NS5 branes on themselves. The NS5 brane is a source in the 
Bianchi identity for $H_3$ (\ref{eom}), but does not affect the equations (\ref{pert-eom}). Therefore, the 
terms which enter the NS5 brane action are not changed, and thus, there is no NS5 self coupling.

\subsection{\large Unequal masses} 

As we have mentioned above, we would like to study a supersymmetric case, where we 
can compute the third term in the potential. We will examine the case when three fermion masses are equal, 
and the fourth is taken to zero. In this limit supersymmetry is restored, and the 3-sphere becomes a very long SO(3) 
invariant ellipsoid.

In (\ref{fermass}) we chose $m_1=m_2=m_3=m$, and $m_4 \ll m$. 
We also assume that the 
ellipsoid is situated in the 3456 plane. This configuration can be generalized to other $SO(3)$ 
invariant configurations by a simple phase rotation by the same angle in the 3-7,4-8 and 5-9 planes. 
The ellipsoid is parametrized  
\beq 
x^6&=& \alpha\  r \cos \theta, \nonumber \\ 
x^3&=&r \sin \theta \cos \theta_1, \nonumber \\ 
x^4&=&r  \sin \theta \sin \theta_1 \sin \phi, \nonumber \\ 
x^5&=&r  \sin \theta \sin \theta_1 \cos \phi,  
\label{par}
\enq 
where $\alpha$ gives the squashing of the ellipsoid, and depends on the ratio $m_4 /m$. Since in general 
the induced fields are not  constant on the ellipsoid, it is more useful to use $\theta,\theta_1 $ and 
$\phi$ as coordinates. The field strength $F_{\theta\theta_1\phi}$ (\ref{f3})  remains unchanged, 
while $C_{\theta\theta_1\phi}$  (\ref{c3}) is multiplied by $\alpha$ and modified by the change in 
(\ref{fermass}). Also
\be
G_{\perp} = r^6 Z^{3/2}  \sin^4 \theta \sin^2 \theta_2 (\alpha^2 \sin^2 \theta + \cos^2 \theta).
\ee

We have observed in chapter 4.2 that the equations of motion of an NS5 brane with large D2 charge 
create a dual field almost equal to the background $-C_{012}$, regardless of $G_{\perp}$,  
so there is no new contribution from the Wess-Zumino term. 
The dominant terms of the potential do not depend on $G_{\perp}$, so they  have the same value as 
before, and they cancel as in the SO(4) invariant case. The terms which before were proportional 
to $ r^6 / A\ $ are now variable on the ellipsoid, and their value is:
\be
V_{\sim r^6} \sim \int_{E^3}{\sqrt{G_{\parallel}} {G_{\perp} \over D_{\theta \phi \alpha }}}. 
\ee
The integral gives:
\be
{-S_{\sim r^6} \over 2 \pi^2 \mu_5 V} = {3 r^6 \over 4 A}{ 3 \alpha^2 + 1  \over 4}. 
\ee
We can interpret $x^6$ as the real part of a complex variable $w$, whose imaginary part is forced to zero. 
This picture fits very well with the intuition coming from reducing the M-theory picture to IIA. Indeed, as 
we go from M-theory to IIA the eleventh dimension $x^{10}$ shrinks to zero. 
We can put the back  phase of $r$ and obtain the potential:
\be 
{-S \over 2 \pi^2 \mu_5 V} = {3 \over 16 g_s^3 A} \left({ 3 |z|^4 |w|^2 + |z|^6 }\right) - {1\over 4g_s^2} {\rm 
Re}(3m w zz\bar z+m_4 z^3 \bar w),
\label{sns}
\ee
where the phase of $w$ does not have any meaning. 
In the limit $m_4 \rightarrow 0$ the theory becomes supersymmetric, and it is possible to complete 
the square in (\ref{sns}) to obtain:
\beq 
{-S \over 2 \pi^2 \mu_5 V} &=& {3 \over 16 g_s^3 A} \left({ 3 |z|^4 |w|^2 + |z|^6 }\right) - {1\over 4 g_s^2} {\rm 
Re}(3m w zz\bar z+m_4 z^3 \bar w) \nonumber \\
&+& { A\over 12 g_s} (3 m^2 |z|^2 + m_4^2 |w|^2).
\label{sns2}
\enq

As we explained, the last term in this potential comes from second order corrections to the bulk $C_{012}$,  
metric and dilaton. Since these second order fields also give the last term in the D4 brane effective potential (\ref{totalact}), we expect the last terms to be similar. Indeed, they have the same dependence on $g_s, m^2, r^2$, and only differ by a numerical factor (1/2). This is not unexpected in view of the different ways the dilaton and the graviton enter the D4 respectively the NS5 brane actions.

The potential has supersymmetric minima at:
\beq
z^2 = {2 A g_s \over 3} m \sqrt{m_4 \over m}, \nonumber \\   
x_6^2 = {2 A g_s \over 3} m \sqrt{m \over m_4},
\label{minima}
\enq
where $A =4 \pi n (\alpha')^{3/2}$.
The ellipsoid  has aspect ratio $\alpha = \sqrt{m \over m_4}$, and degenerates into  a very long line in the supersymmetric limit.
In this limit the NS5 flux on the opposite sides of the ellipsoid adds to zero. Thus, we will just have D2 branes at different positions along $x^6$. Like in the study of D2-D4 polarizations, $x^6$ is a modulus.
A general vacuum configuration will be given by a combination of D4 brane spheres and NS5 brane ellipsoids, with D2 brane
charges $n_i$ and at positions given by (\ref{D4minimum}) or (\ref{minima}).

We  allow $m_4$ to be  small but nonzero, so that we can study the NS5 vacua, and in the 
same time  use
supersymmetry. This can make the supersymmetric ground state slightly metastable (with a 
tunneling rate which becomes infinite as $m_4 \rightarrow 0$ ). However, the picture in which the D2 
branes are polarized into an NS5 brane ellipsoid is valid.

It is interesting to  fully study the case with four fermion masses turned on, which depending on the 
magnitude of the third term may or may not produce polarization into spherical NS5 branes. In M-theory 
this configuration is supersymmetric, and M2 branes are polarized into M5 branes wrapped on three-spheres. As 
we flow into IIA we lose the supersymmetry, but intuitively the branes should still remain polarized. 

\subsection{\large SO(4) invariant non supersymmetric solutions}

The general scalar mass term in the boundary theory is:
\be
m^2(\Phi_3^2+ \Phi_4^2+ \Phi_5^2+\Phi_7^2+\Phi_8^2+\Phi_9^2+\Phi_6^2) + \mu_{ij} \Phi_i \Phi_j,
\label{general}
\ee
where $\mu_{ij}$ are general $7 \times 7$ traceless symmetric matrices. The scalar $\Phi_6$ is different from the other scalars because it belongs to the vector multiplet in the supersymmetric case. From (\ref{sns2}) we see that supersymmetry constrains the last term of (\ref{general}) to be 
\be
 \mu_{ij} \Phi_i \Phi_j = {m^2 \over 6}(\Phi_3^2+ \Phi_4^2+ \Phi_5^2+\Phi_7^2+\Phi_8^2+\Phi_9^2- 6 \Phi_6^2 ).
\label{susymu}
\ee
As we have seen in the case of D4 branes and NS5 branes, this mass term is the last term in the action of a wrapped brane. It appears in the supergravity calculation  through the second equation in (\ref{eom}), where the fields (\ref{nonnorm}) give a contribution:
\be
C^2_{012} \sim  m^2 /g_s, 
\ee
and similar ones to the dilaton and the metric. This determines in principle the last term of the action. Nevertheless, the supergravity equations of
motion satisfied by $C_{012}$ have homogeneous solutions, and as explained in \cite{joe}, the solution is not determined completely by the square of the first order perturbations in (\ref{nonnorm}). These perturbations, which are proportional to $T^3 \wedge T^4$, only determine the mass of the $L=0$ mode. There is another $L=2$ piece, the second term in (\ref{general}), which can be specified arbitrarily in the boundary theory. 
This arbitrariness is not present in the supersymmetric theory (\ref{susymu}). 

Since we want to investigate an SO(4) invariant situation (in the 3456 plane), we set $m_4 = m$ in (\ref{fermass}), and set $\mu_{ij} =0$. The last term of (\ref{sns}) is
\be
  {-S_{r^2}\over 2 \pi^2 \mu_5 V} = { A\over 12 g_s} (3 m^2) r^2 = { A\over 12 g_s} (3 m^2){8 \over 3 \pi}\int_0^{\pi}d \theta \sin^2 \theta  (x_3^2+x_4^2+x_5^2+ x_7^2+x_8^2+x_9^2),
\ee 
where the normalization of the last term is obtained using (\ref{par}).
This is properly interpreted as a combination of an $L=0$ and $L=2$ mode:
\beq
 {-S_{r^2}\over 2 \pi^2 \mu_5 V} &=& { A\over 12 g_s} (3 m^2) {8 \over 3 \pi}\int_0^{\pi} d \theta \sin^2 \theta   \left[{6\over 7} (x_3^2+x_4^2+x_5^2+ x_7^2+x_8^2+x_9^2 + x_6^2) \right.  \nonumber \\
&+& \left.  {1\over 7} (x_3^2+x_4^2+x_5^2+ x_7^2+x_8^2+x_9^2-6 x_6^2) \right].
\enq

The coefficient of the $L=0$ term is given by $T_3 \wedge  T_3$ and is proportional to $3m^2$. Turning on the fourth
fermion mass changes this to $4m^2$. Thus the SO(4) invariant action is given by
\be
{-S \over 2 \pi^2 \mu_5 V} = {3 r^6 \over 4 g_s^3 A} - {m \over  g_s^2} r^4  +\frac{8 A m^2 r^2} {21 g_s }.
\label{SO4action}
\ee
\begin{center}
\begin{figure}[t]
\scalebox{0.7}{\includegraphics{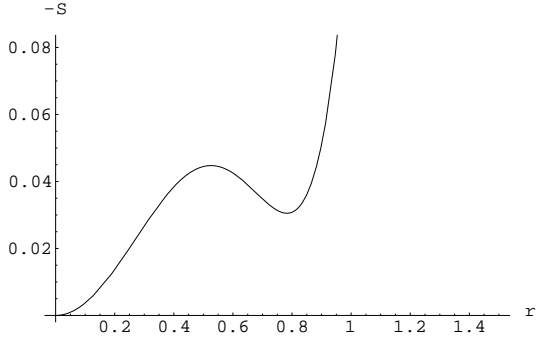}}
\caption{ Effective potential as a function of $r$. The local maximum is at $r=\frac{2\sqrt{g_s A m}}{3}\sqrt{1-\frac{1}{\sqrt{7}}}$ . The local  minimum is at $r_0=\frac{2\sqrt{g_s A m}}{3}\sqrt{1+\frac{1}{\sqrt{7}}}$. }
\end{figure}
\end{center}
This has a local  minimum  at 
\be
r=\frac{2\sqrt{g_s A m}}{3}\sqrt{1+\frac{1}{\sqrt{7}}}.
\label{so(4)min}
\en
We can see that the energy of the NS5 brane at this radius is higher than the one at the origin. It is interesting to explore if other possible D2 brane polarizations have lower energies than this configuration.

We can easily find  the new D4 brane effective potential for the SO(4) invariant case. As we can see from (\ref{fermass}) the second term of (\ref{actionin}) gets modified:
\be
 {\rm Im} (3 z^2 \bar z) \rightarrow {\rm Im}(3 z^2 \bar z + z^3 ). 
\ee
Since this is the term which keeps the brane from collapsing and is the only phase dependent term, we can see that the orientation which maximizes it is $z = r e^{i \pi /6}$.
The last term also changes, in exactly the same way as in the NS5 case. The potential felt by the D4 brane in 
the plane most favorable to polarization is:
\be
{-S \over V} = {2 \mu_4 \over g_s n \alpha'}(r^4 - 5/3 r^3 r_0 + 32/21 r^2 r_0),
\ee
where $r_0 = \pi m n \alpha'$ is the old polarization radius. This potential has only one minimum, at
$r=0$, and no metastable minima at $r \neq 0$. Therefore D2 branes cannot polarize into D4 branes.
This means that the SO(4) invariant theory does not have any Higgs vacua. 
Unfortunately,  we do not know all the vacua of the theory, so we cannot claim based on this that the lowest energy vacuum is confining. 

\subsection{ \large The near shell metric of an NS5 brane wrapped on a 3-ellipsoid}

Let us try to find the metric near the NS5 brane ellipsoidal shell. As in the D4 brane case, we  obtain 
the near shell limit of (\ref{metric}), and we match it with the metric near a flat NS5 brane with D2 
brane flux.

If we denote by $\vec{w}$ the coordinate in the 4-plane of the ellipsoid, and by $y$  the radius in the 3-plane 
transverse to the ellipsoid, we can express the warp factor as:
\be
Z=c \int_{ellipsoid} \frac{\sigma(\vec r )}{((\vec{r}-\vec{w})^2+y^2)^{5/2}},
\en
where $\vec r$ is the coordinate on the ellipsoid, $\sigma(\vec r )$ is the D2 charge density, and  $c$ is 
a normalization constant. 
 The D2 charge is proportional to the absolute value of the 3-form field 
strength $F_{\theta \theta_1 \phi} $ given by (\ref{f3}).
We are interested in the limit of $Z$ as we approach the ellipsoid in two points, as indicated in the Figure 2.   
\begin{center}
\begin{figure}{h}
\scalebox{0.7}{\includegraphics{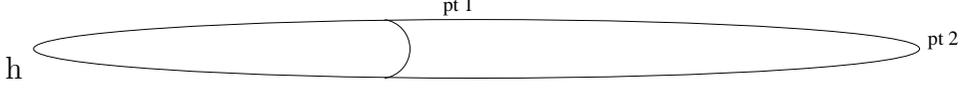}}
\caption{Squashed $4$ ellipsoid in the $(w_1,w_2,w_3,w_4) $ plane. We 
calculate $Z$ at points 1 and 2.}
\end{figure}
\end{center}
\vspace{-1cm}
For point 1 we evaluate $Z$ at  $\vec{w}=(0,w,0,0)$:
\be
Z=\frac{R^5}{\pi}\int \frac{\sin^2{\theta} \sin{\theta_1} d \theta  d \theta_1}{(r_0^2 \alpha^2 
\cos^2{\theta}+r_0^2 \sin^2{\theta}+w^2+y^2-2w r_0 \sin {\theta} \cos {\theta_1})^{5/2}},
\en
where the normalization constant is chosen to match (\ref{2}) far away from the shell. As before 
$\rho^2 \equiv (w-r_0)^2+y^2$. In the $\rho \rightarrow 0$ limit we obtain
\be
Z \approx \frac{2 R^5}{3 \pi \rho^2 r_0^3 \alpha}.
\label{Z1}
\en

For point 2 we evaluate $Z$ at $\vec{w}=(w,0,0,0)$:   
\be 
Z=\frac{2R^5}{\pi } \int_{0}^{\pi} \frac{d \theta \sin^2 \theta }{(\alpha^2 r_0^2 \cos^2 \theta-2 \alpha 
r_0 \cos \theta w+w^2+y^2)^{5/2}}, 
\en 
where we chose normalization so that $Z$ matches (\ref{2}) for large $w$ and $y$. 
This integral can be conveniently rewritten as
\be
Z=\frac{ R^5}{2 (\alpha r_0)^5} \int_{0}^{\pi} \frac{d\theta \sin^2{ \theta}}{((\cos{ \theta}-x-
1)^2+\bar{y}^2)^{5/2}},
\label {zpt2}
\en
where
\beq 
x=(w- \alpha r_0)/(\alpha r_0), \nonumber \\
\bar{y}=y/(\alpha r_0).
\enq
On dimensional grounds we can write  (\ref{zpt2}) in the 
 limit $x \rightarrow 0$ $ \bar{y} \rightarrow 0$ as
\be
\sim \frac{R^5}{(\alpha r_0)^{3/2}} {1 \over \rho^{7/2}}.
\label{zpat2}
\en
where we have ignored a possible homogeneous function of $x$ and $\bar{y}$.
In order to find the tension of flux tubes, and to find where strings drooping from infinity  attach on the ellipsoid, we have to compare (\ref{zpat2}) with (\ref{Z1}). For large $\alpha$, the warp factor near 1 is far bigger than that near 2. Thus, flux tubes attached to 1  will have less energy than those attached to 2. In what follows we will derive the near shell metric in the vicinity of point 1 and in the next chapter calculate the tension of the flux tube attached there.

In \cite{joe}, the near shell metric of D3 branes polarized into IIB NS5 branes was matched with the warped metric in the $\rho \rightarrow 0$ limit. The warp factor is
\be
Z_c \approx \frac{R_c^4}{4 \rho ^2 {r_0}_c^2},
\label{Z1j},
\en
where we put subscript $c$ on all quantities from \cite{joe} to avoid confusion.
In \cite{oz} it was explained how to obtain the metric for IIA NS5 branes with D2 brane charge by T-dualizing  the corresponding IIB NS5 metric with three brane charge.
Keeping all constants of the original IIB solution intact, we obtain\footnote{Compare for example  (38) and (35) of \cite{oz} with (103) of \cite{joe}.}:
\beq
ds^2_{string} &=& \frac{2 {r_0}_c(\rho^2+\rho^2_c)^{1/2}}{R^2_c} \eta_{\mu \nu} dx^{\mu } dx^\nu+  
\frac{R^2_c}{2 {r_0}_c(\rho^2+\rho^2_c)^{1/2}} (dw^idw^i)  \nonumber\\
&+&\frac{R^2_c 
(\rho^2+\rho^2_c)^{1/2}} {2 {r_0}_c} (dw^4 dw^4 +dy dy), \label{ns5near}  \\
e^{2 \phi} &=& g_s^2 \alpha'^2 \frac{ \sqrt{\rho^2+\rho_c^2} }{\rho^2}, \hspace{1cm} { r_0}_c=m_c \pi 
\alpha' g_s N ,  \nonumber \\
\rho_c &=& \frac{2{r_0}_c \alpha'}{R^2_c}, \hspace{2cm} R^4_c=4 \pi g_s N 
\alpha'^2, \nonumber
\enq
where the $i$'s run from 1 to 3. Let us stress that the constants in the above formula do not have any direct physical interpretation in type IIA theory.  We have a metric which matches (\ref{metric}) with  (\ref{Z1j}). What we need is a metric which matches (\ref{metric}) with  (\ref{Z1}). 
The required metric in IIA language is obtained by a formal identification of  $Z$ and $Z_c$ in the large $\rho$ limit.
Then 
\beq 
\pi  g_s N m_c^2=3 \pi r_0^3\alpha/R^5, \hspace{1cm} r_0^2=\frac{8 \pi^2 }{3 \alpha} N\alpha'^{3/2} g_s m. 
\label{identification}
\enq                                                               
We also obtain
\be
 \ \rho_{\rm crossover} \sim r_0 \left (\ \frac{\alpha^3 m \sqrt{\alpha'}}{N \alpha g_s } \right )^{1/4} .           
\ee
The crossover distance is small compared to $r_0$ which makes the near shell approximation valid.
Since the transverse three sphere has size $ \alpha'$ for a single  NS5 branes, string theory corrections are marginal for the  near shell metric \footnote{The action for coincident NS5 branes is not known.}.
This completes the derivation of near shell metric for NS5 brane ellipsoids with D2 flux.
\subsection{\large The near shell metric of an NS5 brane wrapped on a 3-sphere}
If the NS5 brane is polarized into a 3-sphere the warp factor is
\beq
Z=\frac{2R^5}{\pi} \int_{0}^{\pi} \frac{\sin^2 \theta d \theta}{(r_0^2+w^2-2r_0w \cos \theta +y^2)^{5/2}}= \nonumber \\
\frac{R^5}{(r_0^2+w^2+y^2)^{5/2}} \hspace{1mm} {_2F_1} \left ( \frac{5}{4},\frac{7}{4};2;\frac{4r_0^2 w^2}{(r_0^2+w^2+y^2)^2} \right ).
\label{ns5warp}
\enq
Introducing as before the variable $\rho=(w-r_0)^2+y^2$ and working in the $\rho \ra 0$ limit we find
\be
Z \approx \frac{2 R^5 }{3 \pi \rho^2 r_0^3}.
\label{spherez}
\en
Let us note that we could have obtained (\ref{spherez}) from (\ref{Z1}) by simply making the ellipsoid  a sphere ($\alpha=1$). However $r_0$ for the ellipsoid is different than $r_0$ for the sphere.
The metric is given by  (\ref{ns5near}), where instead of  (\ref{identification}) we make the identification
\be
\pi  g_s N m_c^2=3 \pi r_0^3/R^5, \hspace{1cm} r_0=\frac{4\sqrt{g_s N  \pi \alpha'^{3/2} m}}{3}\sqrt{1+\frac{1}{\sqrt{7}}}. 
\label{sphereid}
\en           
We will use this formula in the next section to calculate the tension of flux tubes in this vacuum.

\section{\large Gauge Theory} 
 
In this section we discuss the string theory duals of the gauge theory objects.First  we compute
the tension of flux tubes, and show that a vacuum dual to D2 branes polarized into NS5 branes is confining. 
Then we  give string theory interpretations of  domain walls, condensates, baryon vertices and glueballs. 

\subsection{Flux tubes} 
In the framework of the AdS/CFT duality, there is a correspondence between Wilson lines in the boundary theory and fundamental string world sheets in the bulk. This correspondence was not initially extended to non conformal theories because the geometry is not valid throughout the bulk. For Dp branes with $p<3$ the string world-sheet would probe regions near the branes where supergravity is not valid because of the large dilaton.
Brane polarization solves this problem by smoothing out the geometry. Therefore, we can do concrete computations of flux tube tensions in various vacua.

In a vacuum with  D2 branes polarized into D4 branes, a fundamental string lowered from the boundary  attaches to the D4 brane shell. It costs no energy to move its ends apart on the boundary. Therefore, vacua with only D4 branes are screening. Another way to see this is by noticing that the parallel components of $G_{\mu \nu}$ in (\ref{d4near}) go to zero as we approach the D4 branes.

Let us find the tension of a hanging string in a vacuum with NS5 branes.
Since $Z$ diverges at small distances from the D2-NS5 shell, the string can lower its tension
 by getting closer to the shell. However, at some point the geometry (\ref{metric}) ceases to be valid, and we need to use the near shell metric (\ref{ns5near}). 
In contrast to the D4 case, $G_{\parallel}$ has a minimum value  of $  \pi \alpha' m_c^2 g_sN$. Thus, a flux tube has
a finite tension:
\be
\tau_e=m_c^2 g_s N/2=4 \pi \left ( \frac{2m}{3} \right )^{3/2}\frac{\sqrt{ N g_s/\alpha}}{\alpha'^{1/4}}.
\en
The tension of the flux tube scales as expected with $g_{YM}^2 N$ and vanishes in the limit of a strongly squashed ellipsoid.
The tension of the fundamental string attached to   a spherically wrapped NS5 brane is
\be
\tau_e=\frac{16m^{3/2} \sqrt{g_s N/\pi}}{27\alpha'^{1/4}} \left (1+\frac{1}{\sqrt{7}} \right )^{3/2}.
\en

\subsection{Domain walls}

Since our theory has many discrete vacua, we expect to have domain walls. 
The simplest to understand are the domain walls between vacua containing D4 branes. 
If the two vacua have the same number of D4 branes, they  meet on a 2-sphere
where they exchange the D2 brane charge. The tension of the domain wall is
the bending tension of this configuration.
If the vacua have different numbers of D4 branes, they  meet on a 2-sphere, 
and by charge conservation the appropriate number of D4 branes span the 3-ball 
that has the 2-sphere as the  boundary.

Let us try to understand the domain wall between a vacuum with an $S^2$ brane wrapped 
D4 brane and the one with an NS5 brane wrapped on a 3-ellipsoid. The NS5 brane is extended in the 3456 direction, while the D4 brane is extended in the 789 directions. As they get close they  bend towards each other, but since the D4 is lighter it bends more. By charge conservation at the intersection point there is an NS5 brane with D4 brane charge on it, spanning the squashed ball that has the ellipsoidal boundary. 

There are also other vacua in this theory, characterized by oblique configurations of NS5 branes with D4 flux. They will be studied together with their domain walls in \cite{us-2}.

\subsection{Condensates}
The coefficient of the normalizable mode in the polarized brane background determines the value of gauge theory condensates. The highest normalizable mode (\ref{norm}) decays at infinity as $r^{-7}$. Its coefficient $\bar \alpha$ gives the value of a condensate containing the fermion bilinear and its supersymmetric partners. 
As an example consider the normalizable mode sourced by a D4 spherical shell. Since the D4 brane is a magnetic source for $F_4^1$, the Bianchi identity for $F$ (\ref{eom}) gets modified to:
\be
d F_4 = J_5,
\ee
with
\be
J_5 \sim \alpha'^{3/2} \delta^4(y) \delta(w-r_0) dw \wedge d^4 y,
\ee
where the 2-sphere has radius $r_0$ in the $w$ 3-plane and the 4 transverse coordinates are denoted by $y$.
Without doing the computation we can see by dimensional analysis that
\be 
H_3^{\rm normalizable} \sim {\alpha'^{3/2} r_0^4 \over r^7}\ dw^1 \wedge dw^2 \wedge dw^3.
\ee 
This gives 
\be
<\lambda \lambda> \sim m^4 N^4 \alpha'.
\ee
The exact numerical coefficient of the normalizable mode can be computed through a straightforward generalization of the procedure in \cite{joe}. Nevertheless, since the condensate contains a combination of the fermion bilinear and its supersymmetric partners, the individual condensates are not known. 

It is harder to find the condensates in the confining vacuum. The ellipsoidal geometry complicates things. However, we can estimate the condensates by dimensional analysis
\be
<\lambda \lambda> \sim  1/\alpha ^k (m N)^{5/2} \alpha'^{1/4},
\en
where the power $k>0$ of the squashing ratio $\alpha$ cannot be determined by dimensional analysis alone.
Note that the condensate disappears in the limit of an ellipsoid squashed to a line since the Ramond-Ramond charges on the opposite sides of the ellipsoid cancel.
The condensate corresponding to the NS5 brane wrapped on a three sphere is simply
\be
<\lambda \lambda> \sim  (m N)^{5/2} \alpha'^{1/4}.
\en

\subsection{\large Baryon Vertices.}

It was argued in \cite{witten} that the dual of a baryon vertex in a 4-dimensional $\cN=4$  $SU(N)$ SYM is
a D5 brane wrapped on the five sphere in the $AdS_5 \times S^5$ geometry. In the 3-dimensional $\cN=8$
theory one expects to find the dual of a baryon vertex in terms of a D6 brane of geometry $R \times S^6$. 

At a very high energy the non perturbed picture of the baryon vertex is still valid. We expect this picture to change when the energy becomes comparable to the fermion masses.

In a confining vacuum, as a shrinking D6 brane crosses the NS5 brane shell, a D4 brane is created via the Hanany-Witten effect \cite{nanany}. This D4 brane  fills the ball whose boundary is the NS5 three-sphere (or three-ellipsoid).

A D4 brane ending on an NS5 brane is a source for the 1-form field strength living on the NS5 brane. The D2 branes dissolved in the NS5 brane become $F_3$ flux. The world-volume action of the NS5 brane contains a term \cite{sorokin}:
\be
\int F_3  \wedge B_2 \wedge \cal{F}.
\en 
We can see that in order for this term to be invariant under the gauge transformation $\delta B_2 = d \xi_1$, $N$ fundamental strings should end on the D4-NS5 junction. Therefore this configuration represents a baryon vertex.
The energy of this D4 vertex is
\be
 \mu_4 \int_{B^4} d^4 x e^{-\Phi} G_{induced}^{1/2} =  \frac{\mu_4}{g_s} \int_{B^4} d^4 x  Z^{1/2}.
\en
The integral is divergent near $r_0$ and has to be regularized at $r_0-\rho_c$.
The warped $Z$ is given in terms of the hypergeometric function (\ref{ns5warp}).
To get the leading contribution to the baryon  vertex mass we use the small $\rho$ limit (\ref{spherez}) to obtain:
\beq
\frac{\mu_4 (2R^5/3\pi r_0^3)^{1/2}}{g_s} \int^{r_0-\rho_c}2\pi^2 w^3 d w \frac{1}{w-r_0} \approx \\
 \approx N \left (\frac{m^3 N g_s}{\sqrt{\alpha'}} \right)^{1/4} \ln (g_s N) \left( \frac{1}{3\sqrt{\pi}} \sqrt {1+\frac{1}{\sqrt{7}}} \right )^{3/2}.
\enq
It would be interesting to also understand baryon vertices in the other vacua.

\subsection{\large Instantons.} 

Our candidates for instantons, objects of space time dimension 0,  are either fundamental strings wrapped 
around the D4 two-sphere or D2 branes wrapped on the NS5 three-ellipsoid. None of this configurations is stable in 
string theory: a wrapped string can lower its tension by attaching to the D4 brane and unwrapping. Similarly a wrapped D2 brane can lower its tension by attaching and sliding from the NS5 brane.

\subsection{\large Glueballs and other states.}

A supergravity state localized on a two sphere has transverse momentum $k_m \sim 1/r_0$. Its energy is found by requiring
\be
G_{\mu \nu} k^{\mu}k^{\nu} \sim G_{mn}k^m k^n,
\en
which gives the glueball mass
\be
k_\mu \sim \left (\frac{r_0}{R} \right )^{5/2} \frac{1}{r_0} \sim \frac{ \alpha'^{1/4} m^{3/2} N}{ g_s^{1/2}}.
\en
A typical string state, on the other hand has
\be
G_{\mu \nu} k^{\mu}k^{\nu} \sim 1/\alpha',
\en
which translates into the energy scale 
\be
k_{\mu} \sim \frac{\alpha '^{1/8} m^{5/4} N}{g_s^{1/4}}.
\en
We encounter the same puzzle as in \cite{joe}. There is no sign of gauge theory states with masses given in the previous equations.

\section{\large Flows to M theory.} 
It is interesting to examine what happens to our configurations as we lift them to M-theory. Both the NS5 and the D2 
branes correspond to unwrapped M5 and M2 branes respectively. Thus, the 
D2-NS5 configurations just become the M2-M5 configurations studied in \cite{iosif}.  In the case of D2-D4 configurations, as we increase the radius of the 11'th dimension, the D2 branes remain localized, while
the D4 branes reveal their true nature. They become M5 brane cylinders of geometry $R^3 \times S^2 \times S^1$. As the radius 
of $S^1$ grows beyond the radius of the $S^2$, the M5 brane discovers that it is energetically favorable to be wrapped on a 
3-sphere (or 3-ellipsoid). Thus, there will be a phase transition between these configurations when $g \sqrt{\alpha'} \sim r_0$.

We have not studied oblique configurations  (NS5 branes with D4 flux) in this paper. If they exist, it would be interesting to see what  happens to them in the M-theory limit. Intuitively one may expect them to become either a tilted M5 brane (although this configuration is not a ground state in M theory), or two M5 branes polarized in orthogonal directions.

\section{\large Conclusions and future directions.} 

From a general relativity viewpoint the most important conclusion of this paper is the absence of naked singularities in the gravity duals of a perturbed 3-dimensional gauge theory.

We have shown the existence of supersymmetric  Higgs and non supersymmetric confining vacua in a three-dimensional  Yang Mills theory. We have found their IIA string theory duals in terms of D2 branes polarized into D4 and NS5 branes respectively. The Higgs vacuum of D2-D4 system is expected from the T duality of D3-D5 system studied by Polchinski and Strassler. 

We found a nontrivial realization of a baryon vertex as D4 filling the inside of the NS5 brane 3-sphere. 
Other important results of this paper are the calculation of tensions of flux tubes, an estimate of condensates and discussion of  domain walls in  Yang Mills theory with no supersymmetry.

There are still some unsolved problems. First, it would be interesting to better understand oblique vacua (which may correspond to D2 branes polarized into NS5 branes with D4 charge). Second, the tension of domain walls and their shape could be obtained. Also, the nontrivial connection between the NS5 brane action and the supergravity equations of motion deserves further study. 

The  warping mechanism responsible for the resolution of singularities awaits a fuller understanding. It is the hope of the authors that this paper is a step in that direction.

\section{\large Acknowledgements} 
We are very grateful to Joe Polchinski for suggestions and for reading the manuscript. 
We thank Alex Buchel, and Costin Popescu for numerous discussions. 
We also thank Marina Nudelman for help with  proofreading.
The work of I.B. was supported in part by NSF grant PHY97-22022 and the work of A.N. was supported 
in part by DOE contract DE-FG-03-91ER40618.

\section{\large Appendix} 
Some useful relations between four-tensor spherical harmonics
\beq 
T_4&=&\frac{1}{4!} T_{mnpk} dx^m \wedge dx^n \wedge dx^p \wedge dx^k \ ,\nonumber \\ 
V_{mnpk}&=&\frac{x^q}{r^2} (x^m T_{qnpk}+x^n T_{mqpk}+x^p T_{mnqk}+x^k T_{mnpq})\ , 
\nonumber \\ 
V_4&=&\frac{1}{4!} V_{mnpk} dx^m \wedge dx^n \wedge dx^p \wedge dx^k\ ,  \nonumber \\ 
S_3&=&\frac{1}{3!}T_{mnpk} x^m \wedge dx^n \wedge dx^p \wedge dx^k\ ,  \nonumber \\ 
dS_3&=&4 T_4\ , \ \ \ \   dV_4=-4 d (\ln r) \wedge T_4\ , \ \ \ \  d (\ln r) \wedge S_3=V_4\ ,  \nonumber \\ 
d T_4&=& 0\, \ \ \ \ d(r^p S_3) = r^p (4 T_4+p V_4) \ ,
\enq
where $r^2 = x^m x^m$. The 11 dimensional Hodge duals can be converted into flat metric Hodge duals 
on the 7-dimensional space 
transverse to the D2 branes:
\beq
 {*} F_4=\frac{1}{Z} (*_7 F_4) \wedge  dx^0\wedge dx^1\wedge dx^2 \nonumber \\ 
{*} H_3=\frac{1}{Z^{1/2}}  (*_7H_3) \wedge dx^0\wedge dx^1\wedge dx^2 
\enq

For completeness we also give the formulas from \cite{joe}, which relate 3-tensor spherical harmonics.
\beq
T_3&=&\*  \frac{1}{3!}  T_{mnp} dx^m\wedge dx^n\wedge dx^p, \ \ \ \ S_{ 2}=  \frac{1}{2} T_{mnp} 
x^m dx^n\wedge dx^p\ , \nonumber \\
V_{mnp} &=& \frac{x^q}{r^2} ( x^m T_{qnp} + x^n T_{mqp} + x^pT_{mnq}), \ \ \ \  V_{ 3} = 
\frac{1}{3!}  V_{mnp} dx^m\wedge dx^n\wedge dx^p \ , \nonumber \\
dS_2 &=& 3T_3\ ,\ \ \ \  d(\ln r)\wedge S_2 = V_3\ ,\ \ \ \   dV_{3} = -3 d(\ln r) \wedge T_{3}\ 
,\nonumber\\
dT_{3} &=& 0\ ,\ \ \ \  d(r^p S_2) = r^p(3 T_3 + p V_3)\ .
\enq

where $r^2 = x^m x^m$.

\begin{thebibliography}{1} 
\bibitem{juang} J. Maldacena, Adv. Theor. Math. Phys. 2 (1998) 231, hep-th/9711200
\bibitem{myers} R. Myers, JHEP 9912, 022 (1999), hep-th/9910053 
\bibitem{joe} J. Polchinski and M. Strassler, hep-th/0003136 
\bibitem{magoo} O. Aharony, S. Gubser,J. Maldacena, H. Ooguri, Y. Oz,Phys.Rept.323 (2000) 183,  hep-th/9905111 
\bibitem{seiberg} N. Seiberg, Nucl.Phys.Proc.Suppl. 67 (1998) 158, hep-th/9705117
\bibitem{oz} M. Alishahiha, Y. Oz, M.M. Sheikh-Jabbari, JHEP 9911 (1999) 007, hep-th/9909215 
\bibitem{sorokin} I. Bandos, A. Nurmagambetov, D. Sorokin, hep-th/0003169 
\bibitem{imsy} S. Itzhaki, J. Maldacena, J. Sonnenschein and S. Yankielowicz,Phys.Rev. D58 (1998) 
046004, hep-th/9802042
\bibitem{witten} E. Witten, JHEP 9807 (1998) 006, hep-th/9805112 
\bibitem{nanany} A. Hanany and E. Witten, Nucl. Phys. B492 (1997) 152, hep-th/9611230 
\bibitem{iosif} I. Bena, hep-th/0004142
\bibitem{pst} P. Pasti, D. Sorokin and M. Tonin, Phys.Lett B398 (1997) 41, hep-th/9701037 
\bibitem{ps} M. Perry and J.Schwarz, Nucl.Phys. B395 (1997) 191, hep-th/9611065 
\bibitem{schwarz}J. Schwarz , Phys. Lett B395 (1997) 191, hep-th/9701008 
\bibitem{costin} M. Aganagic, C. Popescu, J. Schwarz, Nucl. Phys. B495 (1997) 99, hep-th/9612080
\bibitem{us-2} I. Bena and  A. Nudelman, hep-th/0006102
%%CITATION=hep-th/0006102;%%
\end{thebibliography}
\end{document}